# FEMTOSECOND LASER WRITTEN DIAMOND WAVEGUIDES: A STEP TOWARDS INTEGRATED PHOTONICS IN THE FAR INFRARED

Vibhav Bharadwaj[1,*], Yuchen Wang[1,2], Toney Teddy Fernandez[1,3], Roberta Ramponi[1,2], Shane M. Eaton[1,2], Gianluca Galzerano[1,2]

[1] *Istituto di Fotonica e Nanotecnologie (IFN) – Consiglio Nazionale delle Ricerche (CNR), Piazza Leonardo da Vinci 32, 20133 Milano, Italy*

[2] *Dipartimento di Fisica - Politecnico di Milano, Piazza Leonardo da Vinci 32, 20133 Milano, Italy*

[3] *Present affiliation: MQ Photonics - Department of Physics and Astronomy, Macquarie University, North Ryde NSW 2109, Australia*

*Corresponding author: vibhavbharadwaj@gmail.com

**Abstract**

A first demonstration and complete characterization of mid-infrared waveguides in single crystal diamond are reported. Waveguides were designed for 2.4 µm and 8.6 µm waveguiding, with their group velocity dispersion analyzed using femtosecond pulses at 2.4 µm wavelength propagated through the waveguide and the bulk substrate. The total dispersion was found to be dominated by the bulk material rather than the waveguide, and was on the range of 275 fs$^2$/mm, demonstrating that femtosecond laser written modifications in diamond introduce negligible perturbations to the pristine material.

**Keywords:** Diamond photonics; Diamond machining; mid- and far-infrared photonics; Laser materials processing; Femtosecond phenomena;

## 1. Introduction

Diamond is a supreme material, considering its excellent physical, chemical and optical properties due to its strong and dense tetrahedral covalent bonds that are highly structured to form a cubic lattice. The optical transparency offered by diamond is one of the widest in naturally occurring materials, extending from the UV region (~225 nm) to terahertz (THz) frequencies and even beyond into the microwave region (~8000 µm). Additionally, diamond has low group velocity dispersion from the mid-IR region up to THz region[1], thus projecting itself as a suitable platform for integrated photonics over a broad wavelength range. The mid-IR region of the electromagnetic spectrum, from 2 to 20 µm, has been shown as a promising wavelength range for molecular spectroscopy and sensing[2,3] enabling the identification and classification of analytes with ultra-high precision and rapidness with non-ionizing and non-invasive/minimally invasive techniques. Additionally, the earth's atmosphere shows broad transmission windows in this spectral region, allowing the use of mid-IR wavelengths for remote sensing and optical radar[4,5] as well as astrophotonic applications[6,7]. A suitable platform is crucial for the technological development of integrated mid-IR photonics. Materials like chalcogenide glasses[8], fluoride glasses[9], zinc selenide and sulfide[10], silver halide[11] and tellurium halides[12] have been researched upon due to their wide transparency in the mid-IR wavelengths.

Recently we have reported the fabrication of optical waveguides in bulk single-crystal diamond using femtosecond laser writing technology, which has opened up the prospect of a three-dimensional, integrated quantum photonics platform[13]. Femtosecond laser writing of nitrogen vacancies (NV) into diamond to produce optically coherent NV⁻ color centers at desired locations[14-16] and laser written optical waveguides in diamond that could be used to excite and collect light from those single NVs[16] were also demonstrated recently. Wavelength selective reflective elements such as Bragg waveguides have also been laser-fabricated in the bulk of diamond allowing the ability to create cavities for Raman laser applications in the mid-IR wavelengths[17]. In addition, diamond being a bio-inert material is a suitable substrate for in-situ bio-chemical sensing applications[18]. The availability of 3D laser written diamond waveguides would provide the strong backbone to extend its usage also in the far infrared and the THz region for an all-optical device.

In this paper, we fabricate mid-IR waveguides in single crystal diamond for the first time. The laser written waveguides were designed to operate at 2.4 µm and 8.6 µm wavelengths. Their single mode operation and loss characterization were carried out and then using a Kerr-lens-mode-locked (KLM) Cr:ZnSe laser ultrashort pulses, the waveguides were characterized for their group velocity dispersion performance.

2. **Experimental fabrication and characterization**

Polished 5 mm × 5 mm × 0.5 mm synthetic single-crystal diamond samples (type II, optical grade with nitrogen impurities ~100 part per billion) were acquired from MB Optics. The femtosecond laser used for waveguide writing in diamond was a regeneratively amplified Yb:KGW system (Pharos, Light Conversion) with 230-fs pulse duration at 515-nm wavelength (frequency doubled) and a repetition rate of 500 kHz, focused with a 1.25-NA oil immersion lens (RMS100X-O 100 × Olympus Plan Achromat Oil Immersion Objective). Laser-inscribed structures were characterized for their morphology using white-light optical microscopy in transmission mode with 10× and 40× magnification objectives (Eclipse ME600, Nikon). The structures were formed at 50 µm depth, as measured from the surface to the midpoint of the vertically elongated tracks.

Type II waveguides were designed and fabricated in which two modification lines were laser written with a suitable spacing that enabled light to be guided between them. Total average power of the laser was varied between 75 – 100 mW. The separation between the modification lines was 18 – 35 µm. In all cases the scan speed was 0.5 mm/s. The waveguide performance and their losses were tested for two sets of laser wavelengths in the mid-IR regime: at 2.4 µm using a home built Cr:ZnSe laser[19] and at 8.6 µm by means of a continuous wave (CW) quantum cascade laser (QCL). Figure 1 shows the experimental setup for the diamond waveguide characterization.

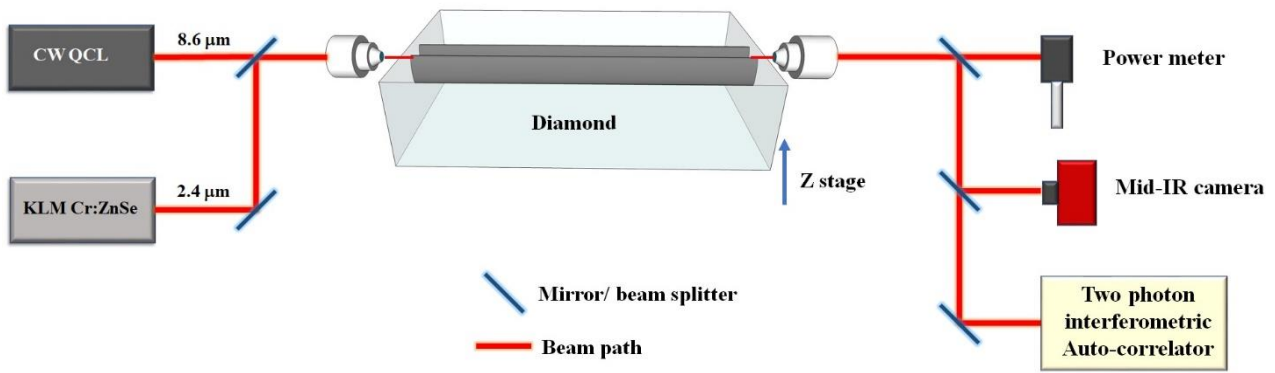

**Figure 1:** Experimental setup for diamond waveguide characterization. CW QCL: continuous-wave quantum cascade laser; KLM Cr:ZnSe: Kerr-lens-mode-locked Cr:ZnSe laser.

At 2.4 µm wavelength the waveguide showed single mode guidance for type II track separations of 30 – 45 µm, and powers between 75 – 100 mW. The best waveguide showed propagation losses of 6 dB/cm for 30 µm separation and 75 mW average power. The optical microscope image of the waveguide is shown in Fig. 2. The structure had dimensions of 6 µm × 23 µm with the strong vertical elongation due to spherical aberration. The optical mode was profiled using a mid-IR camera (WinCamD-FIR2-16-HR, DataRay).

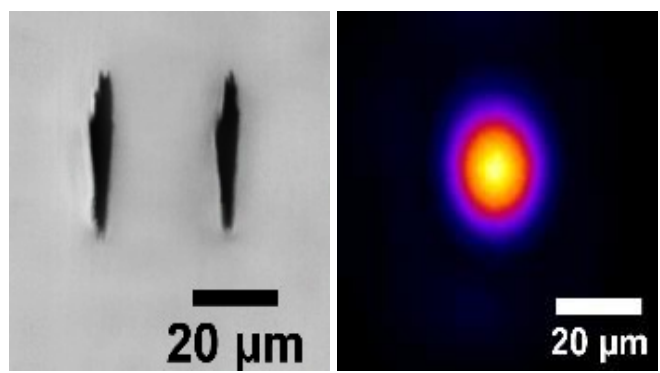

**Figure 2:** Optical microscope image of the transverse view of the Type II waveguide along with the guided mode at 2.4 µm wavelength.

The losses were quite similar at 8.6 µm wavelength albeit the mode sizes were larger for the single mode waveguides. The best transmission was obtained for the waveguide fabricated with 85 mW average power and 40 µm track separation. The cross sectional microscope image of the waveguide and its guided mode are shown in Fig. 3. It was previously demonstrated that stresses were responsible for the waveguiding behavior for visible-guiding waveguides, having a smaller track separation of 13 µm[19]. The mechanisms responsible for waveguiding in diamond in the mid-IR for the larger track separations of 30-40 µm reported here are still under investigation.

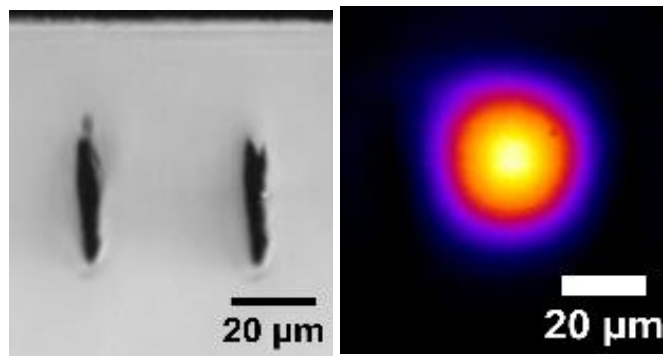

**Figure 3:** Optical microscope image of the transverse view of the Type II waveguide along with the guided mode at 8.6 µm wavelength.

Thanks to the availability of a KLM Cr:ZnSe laser that generates pulse trains at 2.4 µm with of pulse duration of few-optical cycles[20], the group delay dispersion characteristics of the diamond waveguides were studied by coupling transform-limited pulses and measuring the output pulse duration and chirp using an interferometric autocorrelation based on two-photon absorption in an InGaAs photodiode (cutoff wavelength 1.7 µm) as shown in Fig. 1. This chirp is due to a collective contribution from the material and the waveguide dispersion.

The same pulse was sent through the bulk crystal of the same length to measure the dispersion contribution from the material and thereby deducing the waveguide dispersion. The interferometric autocorrelations measured at the waveguide input (a) and output (b), and at the diamond crystal output after propagation through the bulk crystal (c) are shown in Fig. 4. The interferometric autocorrelation of the input pulse train, Fig. 4(a), is characterized by a peak-to-background ratio 8:1 and a coherence peak duration of 80 fs and a negligible pulse chirp, corresponding to a transform limited pulse duration of 52 fs. When the pulses propagate through the diamond crystal, Fig. 4(b) and 4(c), the corresponding interferometric autocorrelations show a slightly shorter coherence peak longer pulse durations and base line distortions, clear indications that the pulses are now chirped. In the case of propagation in the bulk diamond crystal, the intensity autocorrelation trace, blue curve in Fig. 4(b), is characterized by a duration of 116 fs, with a group delay dispersion of |1375| fs$^2$, corresponding to a material group velocity dispersion of |275| fs$^2$/mm. When the pulse train propagated through the diamond waveguide, Fig. 4(c), the duration of the intensity autocorrelation (green curve) is 99 fs, slightly less compared to bulk propagation. The group velocity dispersion is |190| fs$^2$/mm for the waveguide. Femtosecond laser inscription in diamond could induce the of graphitic phases or other allotropes of carbon which can produce undesired effects in the waveguiding operation. The negligible group velocity dispersion arising from the waveguide structure indicates that those detrimental factors are minimal and do not produce any significant impact on the waveguide based applications of diamond.

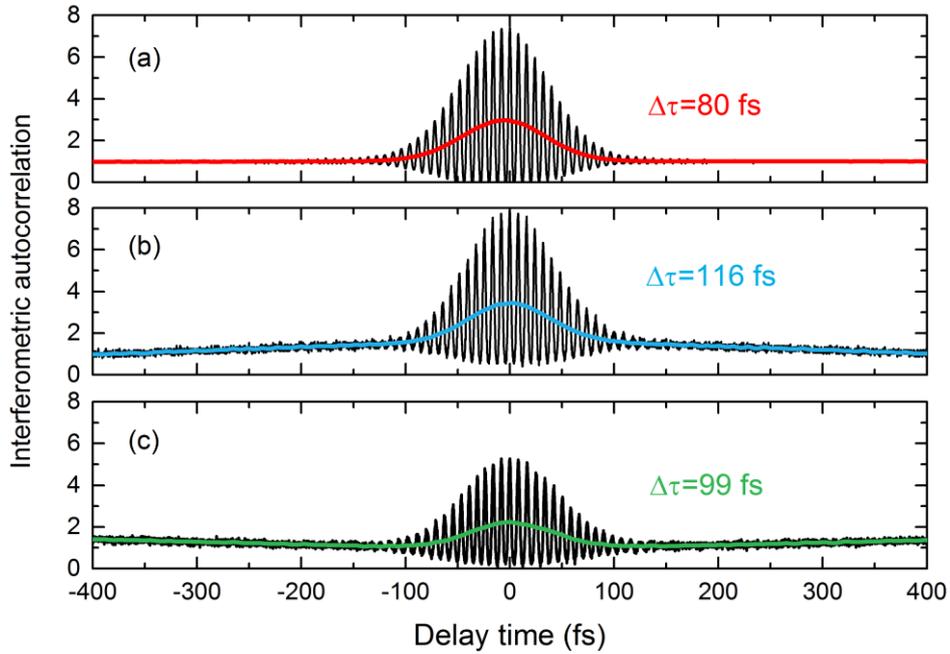

**Figure 4:** Interferometric autocorrelations of the 150-MHz pulse trains generated by the KLM Cr:ZnSe at the input facet of the diamond crystal (a), at the waveguide output (b), and after propagation through bulk diamond crystal (c). The red, blue, and green curves in the autocorrelations correspond to the intensity autocorrelation profile.

3. Conclusion

In this work, we show the first fabrication and characterization of mid-IR waveguides in single crystal diamond. Using femtosecond laser writing, we have demonstrated photonic devices operational at 2.4 µm and 8.6 µm wavelengths, with promise to scale up to longer wavelengths in the far-IR. The group velocity dispersion of the laser-formed waveguides was examined by comparing the interferometric autocorrelations of femtosecond pulses at 2.4 µm propagated through the waveguide and the bulk substrate. Negligible waveguide contribution to the group velocity dispersion to the ultrashort pulses was observed, implying that the total dispersion is dominated by the diamond bulk material, which is in the range of 275 fs$^2$/mm. Supplemented by the 3D femtosecond laser writing technique, and combined with the low group velocity dispersion, diamond photonic devices have great potential to synergize with ultrashort pulses in the mid infrared fingerprint region for on-chip high resolution vibrational spectroscopy and Raman spectroscopy.

**References**

1. H. R. Phillip and E. A. Taft. Kramers-Kronig Analysis of Reflectance Data for Diamond, Phys. Rev. 136, A1445-A1448 (1964).

2. Kosterev, A. *et al.* Application of quantum cascade lasers to trace gas analysis. *Appl. Phys. B* 90, 165–176 (2008).

3. Lucas, P. *et al.* Infrared biosensors using hydrophobic chalcogenide fibers sensitized with live cells. *Sensors Actuators, B Chem.* 119, 355–362 (2006).

4. Walsh, B. *et al.* Mid infrared lasers for remote sensing applications. *J. Lumin.* 169, 400–405 (2016).

5. Hanson, F. *et al.* Single-frequency mid-infrared optical parametric oscillator source for coherent laser radar. *Opt. Lett.* 26, 1794 (2001).

6. Arriola, A. *et al.* Mid-infrared astrophotonics: study of ultrafast laser induced index change in compatible materials. *Opt. Mater. Express* 7, 698 (2017).

7. Arriola, A. *et al.* Ultrafast laser inscription of mid-IR directional couplers for stellar interferometry. *Opt. Lett.* 39, 4820–2 (2014).

8. Tsay, C. *et al.* Solution-processed chalcogenide glass for integrated single-mode mid-infrared waveguides. *Opt. Express* 18, 26744 (2010).

9. Saad, M. Fluoride Glasses and Fiber for Mid-IR Applications. *IEEE Photonics Society Summer Topical Meeting Series 55–56* (IEEE, 2014). doi:10.1109/SUM.2014.39

10. Adams, J. J. *et al.* 40–45-µm lasing of Fe:ZnSe below 180 K, a new mid-infrared laser

material. *Opt. Lett.* 24, 1720 (1999).

11. Grille, R. *et al.* Single mode mid-infrared silver halide asymmetric flat waveguide obtained from crystal extrusion. *Opt. Express 17*, 12516 (2009).

12. Le Neindre, L. *et al.* Tellurium halide optical fibers. *J. Non. Cryst. Solids* 242, 99–103 (1998).

13. Sotillo, B. *et al.* Diamond photonics platform enabled by femtosecond laser writing. *Sci. Rep.* 6, (2016).

14. Chen, Y. C. *et al.* Laser writing of coherent colour centres in diamond. *Nat. Photonics* 11, 77–80 (2017).

15. Sotillo, B. *et al.* Visible to infrared diamond photonics enabled by focused femtosecond laser pulses. *Micromachines* 8, (2017).

16. Hadden, J. P. *et al.* Integrated waveguides and deterministically positioned nitrogen vacancy centers in diamond created by femtosecond laser writing *arXiv*:1701.05885 (2017).

17. Bharadwaj, V. *et al.* Femtosecond laser inscription of Bragg grating waveguides in bulk diamond. *Opt. Lett.* 42, 3451 (2017).

18. Jedrkiewicz, O. *et al.* Pulsed Bessel beam-induced microchannels on a diamond surface for versatile microfluidic and sensing applications. *Opt. Mater. Express* 7, 1962 (2017).

19. Sotillo, B. *et al.* Polarized micro-Raman studies of femtosecond laser written stress-induced optical waveguides in diamond. *Appl. Phys. Lett.* 112, 031109 (2018)

20. Wang, Y. *et al.* 47-fs Kerr-lens mode-locked Cr:ZnSe laser with high spectral purity. *Opt. Express* 25, 25193 (2017).